\documentclass[preprint,letterpaper]{revtex4}

\usepackage{graphicx}
\begin{document}

\preprint{CDF note 5768}
\preprint{D\O\ note 3918}
\preprint{BUHEP-01-30}
\preprint{NUHEP-EXP/01-054}
\preprint{PITHA 01/10}

\title{Testing the Standard Model at the Fermilab Tevatron
\footnote{contribution to the proceedings of the Snowmass Workshop 2001}}

\author{Martin Gr\"unewald}
\email[]{Martin.Grunewald@cern.ch}
\affiliation{RWTH Aachen}
\author{Ulrich Heintz}
\email[]{heintz@bu.edu}
\author{Meenakshi Narain}
\email[]{narain@bu.edu}
\affiliation{Boston University}
\author{Michael Schmitt}
\email[]{schmittm@lotus.phys.nwu.edu}
\affiliation{Northwestern University}

\date{November 16, 2001}

\maketitle

For Run~2a, the Tevatron will deliver 2 fb$^{-1}$ at $\sqrt{s}$ by 2004 to the upgraded CDF and D\O\ detectors, increasing the data from Run~1 twentyfold. Run~2b, inspired by Snowmass 96 and the TeV2000 workshop\cite{TeV2000}, aims for 15 fb$^{-1}$ by 2007, before the LHC begins to do physics. There are discussions of further upgrades to accumulate 30 fb$^{-1}$ by 2008 or 2009. 

\section{$W$ boson mass}

The measurement of the $W$ boson mass from Tevatron Run~1 data achieved a precision of 68 MeV\cite{Wmass}.
A variety of methods can be used to measure the $W$ boson mass with different tradeoffs between statistical and systematic uncertainties. These include fits of the transverse mass and lepton $p_T$ spectra to templates from Monte Carlo simulations. Most systematics, such as the detector calibration and the recoil model, are driven by the number of $Z$ boson decays observed\cite{FNAL_QCDWZ}. Measurements of the charge asymmetry in $W\to\ell\nu$ decays will help constrain the parton distribution functions. New QED calculations will reduce theoretical uncertainties. In the ratio method, developed by D\O\cite{D0_Wmass_ratio}, the $Z$ boson data are rescaled to fit the $W$ boson data. This reduces most experimental and some theoretical uncertainties at the cost of statistical sensitivity. In all cases, systematic uncertainties are expected to dominate. Since the main systematics differ, these methods can be used to check the results for consistency at the 10 MeV level.

Table \ref{P1WG1_schmitt_0713tab1} shows the expected precision of the $W$ mass measurement from the transverse mass fit, extrapolated from the Run~1b measurement by D\O\cite{D0_Wmass_1B_PRD}. The calorimeter scale and linearity assume constraints from $Z$ data only, not the $J/\psi$ and $\pi^0$ data used in Run~1. By about 30 fb$^{-1}$, the determination of the energy resolution will be systematically limited by the uncertainty in the width of the $Z$ boson. The uncertainty due to electron removal was conservatively assumed to decrease only by half.
Table \ref{P1WG1_schmitt_0713tab1} also shows an extrapolation of the uncertainty for the ratio method from Run~1 results by D\O. The systematic uncertainty for this method is smaller than for the transverse mass fit and it may well be the best for high integrated luminosities.
We conclude that the $W$ boson mass will be measured at the end of Run~2 to a precision of 15 MeV, perhaps even 10 MeV, combining the results from both experiments, using several methods and the $W\to e\nu$ and $W\to\mu\nu$ channels.

\begin{table}[htbp]
\caption{Projected uncertainties in MeV of the $W$ boson mass measurement using the transverse mass fit (left) and the ratio method (right) for $W\to e\nu$ decays.}
\label{P1WG1_schmitt_0713tab1}
\centerline{\hbox to 3in{
\begin{tabular}{l|r|rrr}
$\int{\cal L}dt$ (fb$^{-1}$) & 0.08 & 2 & 15 & 30 \\
\hline\hline
statistical uncertainty & 96 & 19 & 7 & 5 \\
\hline\hline
production/decay model 	& 30 & 14 & 13 & 13 \\
backgrounds 		& 11 & 2 & 1 & 1 \\
detector model 		& 57 & 13 & 8 & 8 \\
\hline
total systematic	& 66 & 19 & 16 & 15 \\
\hline\hline
total uncertainty 	& 116 & 27 & 17 & 16 \\
\end{tabular}
}\hbox to 3in{
\begin{tabular}{l|r|rrr}
$\int{\cal L}dt$ (fb$^{-1}$) & 0.08 & 2 & 15 & 30 \\
\hline\hline
statistical uncertainty & 211 & 44 & 16 & 11 \\
\hline
total systematic	& 50 & 10 & 4 & 3 \\
\hline\hline
total uncertainty 	& 217 & 44 & 16 & 12 \\
\end{tabular}
}}
\end{table}

\section{Top quark mass}

In Run~1, the top quark mass was measured to $\approx5$ GeV\cite{CDF_topmass,D0_topmass_lj_PRD}. For Run~2, $t\overline t$ data samples will be large enough to allow a double $b$-tag. For 15 fb$^{-1}$, per experiment 3200 double-tagged single-lepton and 1200 untagged dilepton events are expected.

The main systematic uncertainty for the top quark mass measurement is the jet energy scale. Using Run~1 methods, this uncertainy cannot be reduced below a couple of GeV. However, both experiments plan to use $p\overline p\to Z\to b\overline b$ events, which will help set the energy scale to a precision of about half a GeV\cite{Heintz_Zbb}. In addition, the hadronically decaying $W$ in single-lepton $t\overline t$ events provides an independent calibration point\cite{Snowmass96_top}.

The next most important systematic uncertainty is modeling of gluon radiation in the initial or final state of $t\overline t$ events. In Run~1, this uncertainty was estimated mainly by comparing predictions of different event generators. In Run~2, the modeling of jet activity in top quark events can be constrained better by comparing double-tagged events with simulations. We estimate this uncertainty to be about 1~GeV, and expect it to decrease only slightly with increasing integrated luminosity. 
Other systematics will scale inversely with the square root of the integrated luminosity.   

We extrapolate the uncertainty on the top mass based on Run~1 D\O\ results in the single-lepton channel\cite{D0_topmass_lj_PRD} in Table \ref{P1WG1_schmitt_0713tab3}. We take the higher cross section at $\sqrt{s}=2$ TeV in account and we assume double $b$-tagging with an efficiency per $b$-jet of 65\%. Double $b$-tagging will essentially eliminate the uncertainty due to $W$+jets background.
The uncertainty in the dilepton channel\cite{D0_topmass_ll_PRD} is also extrapolated in Table \ref{P1WG1_schmitt_0713tab3}. No $b$-tagging is assumed here.
For each channel and each experiment, a precision of abut 1.2 GeV is projected. By combining both channels and experiments an overall precision close to 1 GeV should be achievable.

\begin{table}[htbp]
\caption{Projected uncertainties in GeV of the top quark mass measurement in the single-lepton channel (left) and dilepton channel (right).}
\label{P1WG1_schmitt_0713tab3}
\centerline{\hbox to 3in{
\begin{tabular}{l|r|rrr}
$\int{\cal L}dt$ (fb$^{-1}$) & 0.1 & 2 & 15 & 30 \\
\hline\hline
statistical uncertainty & 5.6 & 1.7 & 0.63 & 0.44 \\
\hline\hline
jet scale ($W\to q\overline q$) & 4.2 & 1.8 & 0.64 & 0.45 \\
jet scale ($Z\to b\overline b$) & --- & 0.53 & 0.19 & 0.14 \\
MC model (gluon radiation) & 1.9 & 1.1 & 0.97 & 0.96 \\
event pile-up 		& 1.6 & 0.49 & 0.18 & 0.13 \\
$W$+jets background	& 2.5 & 0 & 0 & 0 \\
$b$-tag			& 0.4 & 0 & 0 & 0 \\
\hline
total systematic	& 5.5 & 2.1 & 1.2 & 1.1 \\
\hline\hline
total uncertainty 	& 7.8 & 2.7 & 1.3 & 1.2 \\
\end{tabular}
}\hspace{0.25in}\hbox to 3in{
\begin{tabular}{l|r|rrr}
$\int{\cal L}dt$ (fb$^{-1}$) & 0.1 & 2 & 15 & 30 \\
\hline\hline
statistical uncertainty & 12.3 & 2.4 & 0.87 & 0.62 \\
\hline\hline
jet scale		& 2.0 & 0.88 & 0.32 & 0.23 \\
MC model 		& 2.3 & 1.0 & 0.96 & 0.96 \\
event pile-up 		& 1.4 & 0.27 & 0.10 & 0.07 \\
background		& 0.9 & 0.17 & 0.06 & 0.05 \\
method			& 0.9 & 0.17 & 0.06 & 0.05 \\
\hline
total systematic	& 3.6 & 1.4 & 1.0 & 1.0 \\
\hline\hline
total uncertainty 	& 12.8 & 2.8 & 1.3 & 1.2 \\
\end{tabular}
}}
\end{table}

\section{Forward-Backward Asymmetry}

The forward-backward asymmetry $A_{FB}$ in the process $u\overline u+d\overline d\to Z\to \ell^+\ell^-$, measured near the $Z$ pole, gives a value of the weak mixing angle $\sin^2\theta_w$. CDF published $A_{FB}=0.070\pm0.015(\mbox{stat})\pm0.004(\mbox{syst})$ based on data from Run~1\cite{CDF_ZAFB_1B}. The statistical uncertainty scales to 0.0016 for 10 fb$^{-1}$ and to 0.0009 for 30 fb$^{-1}$. The most important systematic uncertainty arises from the parton distribution functions. These can be constrained by the charge asymmetry in $W$ decays. A theoretical uncertainty arises from the limited rapidity coverage and the $p_T$ distribution of the $Z$. Both the rapidity and $p_T$ distribution of the $Z$ will be measured and this uncertainty will be reduced. It is expected that the statistical uncertainty will dominate all systematics\cite{FNAL_QCDWZ}.

Combining the electron and muon channels from both experiments leads to a projection for the precision of the $\sin^2\theta_w$ measurement of 0.00028 for 10 fb$^{-1}$ and 0.00016 for 30 fb$^{-1}$\cite{FNAL_QCDWZ}, comparable to the current world average\cite{PDG}. This should clarify the 3.5$\sigma$ discrepancy between $\sin^2\theta_w$ from $A_\ell$ measured at SLD\cite{SLD_ALR} and $A_{FB}^{0,b}$, measured at LEP\cite{LEPEWWG}.

\section{Constraints on the Higgs boson mass}

A significant uncertainty in inferring constraints on the Higgs boson mass from precision electroweak measurements arises from $\Delta\alpha_{had}^{(5)}\left(M_Z^2\right)$, the contribution of light quarks to the running of $\alpha_{EM}$. Its calculation employs $R^{had}(s)=\sigma(e^+e^-\to\mbox{hadrons})/\sigma(e^+e^-\to\mu^+\mu^-)$ from low energy data. The LEP electroweak working group uses a data-driven value\cite{Burkhardt} and also considers a more theory-guided value\cite{Martin}. BES has recently improved the precision of $R^{had}(s)$ to 7\% for $2<\sqrt{s}<5$ GeV\cite{BES_R}. There are plans to measure $R^{had}(s)$ more precisely at CLEO-c\cite{Gibbons_CLEO-c} in the next six years. We assume that $\Delta\alpha_{had}^{(5)}\left(M_Z^2\right)$ will be known to $10^{-4}$, a value already achieved in a recent theory-driven determination\cite{Yndurian}. 

\begin{figure}
\resizebox{5in}{!}{\includegraphics{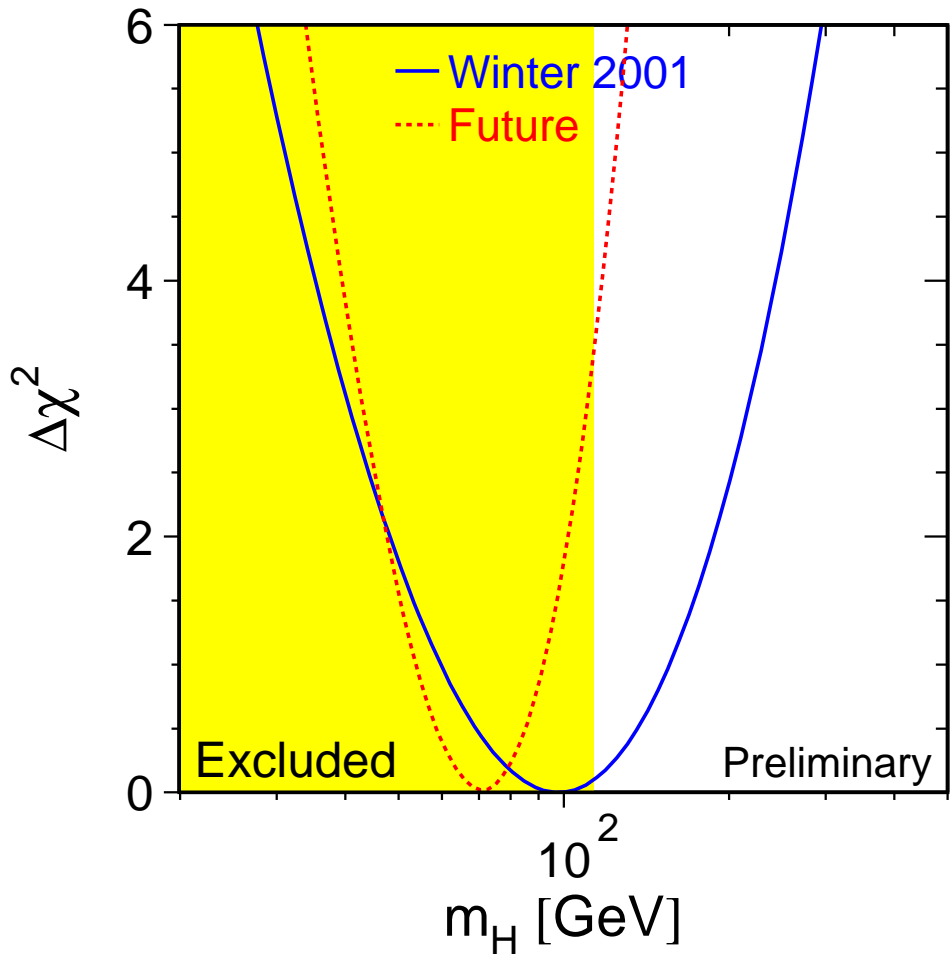}}%
\caption{The $\chi^2$ curve as a function of the Higgs boson mass from the globalk electroweak fit using the projected reduced uncertainties listed in the text.}
\label{P1WG1_schmitt_0713fig1}
\end{figure}

To estimate the effect of the improvements in the measurements described so far, we take the current central values and shrink the uncertainties to 20 MeV for the $W$ mass, 1 GeV for the top mass, and $10^{-4}$ for $\Delta\alpha_{had}^{(5)}\left(M_Z^2\right)$. We are not using the $A_{FB}$ measurement here, since we do not understand the systematics well enough. We then repeat the global electroweak fit. Figure \ref{P1WG1_schmitt_0713fig1} shows the resulting $\chi^2$ as a function of the Higgs boson mass compared to the results from the global fit with the winter 2001 values\cite{LEPEWWG}.

If there is no observation of the Higgs boson in Run~2, the Tevatron can exclude standard model Higgs boson masses below about 185 GeV at the 95\% confidence level\cite{FNAL_SUSYHiggs}. Putting the tight indirect constraints together with the direct lower limit on the Higgs boson mass could severely challenge the standard model. The degree of inconsistency between these data will depend on how much the central values shift. To maximize the potential of the Tevatron to push this test of the standard model, a goal of 30 fb$^{-1}$ of integrated luminosity for Run~2 would be crucial.

\begin{acknowledgments}
This work was in part supported by the U.S. Department of Energy. U. Heintz is an A. P. Sloan Fellow and a Cottrell Scholar of Research Corporation. 
\end{acknowledgments}

\end{document}